\documentclass[aps, nofootinbib]{revtex4}

\usepackage{amsmath}
\usepackage{amssymb}
\usepackage[figuresright]{rotating}
\usepackage{color}
\usepackage{graphicx}
\usepackage{subfigure}

\renewcommand{\v}[1]{\ensuremath{\mathbf{#1}}} 
\newcommand{\gv}[1]{\ensuremath{\mbox{\boldmath$ #1 $}}} 
\newcommand{\abs}[1]{\left| #1 \right|} 
 
\newcommand{\ket}[1]{\left| #1 \right>} 
\newcommand{\bra}[1]{\left< #1 \right|} 

\begin{document}

\title{Does quantum entanglement in DNA synchronize the catalytic centers of type II restriction endonucleases?}

\author{P. Kurian$^{1,2,3}$, G. Dunston$^{3,4}$, J. Lindesay$^{1,2}$}
\address{1. Department of Physics and Astronomy, Howard University, Washington, District of Columbia 20059 \\ 2. Computational Physics Laboratory, Howard University, Washington, District of Columbia 20059 \\ 3. National Human Genome Center, Howard University College of Medicine, Washington, District of Columbia 20059 \\ 4. Department of Microbiology, Howard University College of Medicine, Washington, District of Columbia 20059}

\begin{abstract}
Several living systems have been examined for their exhibition of macroscopic quantum effects, showcasing biology's apparent optimization of structure and function for quantum behavior. Prevalent in lower organisms with analogues in eukaryotes, type II restriction endonucleases are the largest class of restriction enzymes. Orthodox type II endonucleases recognize four-to-eight base pair sequences of palindromic DNA, cut both strands symmetrically, and act without an external metabolite such as ATP. While it is known that these enzymes induce strand breaks by nucleophilic attack on opposing phosphodiester bonds of the DNA helix, what remains unclear is the mechanism by which cutting occurs in concert at the catalytic centers. Previous studies indicate the primacy of intimate DNA contacts made by the specifically bound enzyme in coordinating the two synchronized cuts. We propose that collective electronic behavior in the DNA helix generates coherent oscillations, quantized through boundary conditions imposed by the endonuclease, that provide the energy required to break two phosphodiester bonds. Such quanta may be preserved in the presence of thermal noise and electromagnetic interference through the specific complex's exclusion of water and ions surrounding the helix, with the enzyme serving as a decoherence shield. Clamping energy imparted by the decoherence shield is comparable with zero-point modes of the dipole-dipole oscillations in the DNA recognition sequence. The palindromic mirror symmetry of this sequence should conserve parity during the process. Experimental data corroborate that symmetric bond-breaking ceases when the symmetry of the endonuclease complex is violated, or when environmental parameters are perturbed far from biological optima. Persistent correlation between states in DNA sequence across spatial separations of any length---a characteristic signature of quantum entanglement---may be explained by such a physical mechanism.
\end{abstract}

\maketitle

\section{Introduction}
Macroscopic quantum effects in biological systems have been studied with verve in recent years, as researchers have sought fundamental explanations for diverse phenomena in bacteria \cite{Sarovar}, plants \cite{Sarovar}, birds \cite{Gauger}, flies \cite{Franco}, and humans \cite{Brookes}. Ensconced in thermally turbulent aqueous environments, biology appears to have found mechanisms to optimize structure and function for quantum behavior. How does biology exploit quantum effects when human-made macroscopic quantum states are frustrated by the stringent requirements of extreme cold, vacuum, and isolation from the electromagnetic environment? 

The revolution in genetic sequencing and engineering has been made possible through the use of restriction endonucleases. Originally isolated from bacteria, these enzymes cut DNA at recognition sequences with high specificity, thereby assuring a consistent final product. Each endonuclease has been named using a system loosely based on the bacterial genus, species, and strain from which the enzyme is derived: the first identified endonuclease in the RY13 strain of \textit{Escherichia coli} is \textit{Eco}RI, and the fifth endonuclease extracted from the same strain is \textit{Eco}RV. The body of literature on the structure and catalytic mechanisms of these molecular workhorses is substantial and has been reviewed in multiple instances \cite{Modrich, Jeltsch, Jen-Jacobson, Pingoud1, PingoudV, Roberts, PingoudV2, Type2}.

Orthodox type II restriction endonucleases are unique among these enzymes, as they are comprised of two identical subunits and do not require the assistance of nucleotide triphosphates (e.g., ATP, GTP) or S-adenosyl-L-methionine (AdoMet) for their activity. These enzymes bear a curious twofold rotational symmetry that is reflected in the palindromic symmetry of the double-stranded DNA substrates to which they bind. Orthodox type II restriction endonucleases cleave DNA in a manner that preserves this symmetry. Before cutting, these enzymes rapidly scan the DNA strand in a nonspecifically bound manner by facilitated diffusion \cite{Type2} searching for their recognition sequences, which are between four and eight base pairs (bp) in length. Upon recognition sequence binding, conformational changes take place in the enzyme and DNA, which release water and charge-countering ions from the protein-DNA interface.

How sequence recognition proceeds to catalysis is perhaps the least understood aspect of the enzymology. It is known that concerted cutting of both DNA strands requires intersubunit correlations to synchronize the two catalytic centers of \textit{Eco}RV-like type II endonucleases \cite{Pingoud3}. One study \cite{Pingoud2} revealed that the \textit{Eco}RV DNA-binding domains cannot function independently of each other, and that only with asymmetric modifications can an \textit{Eco}RV mutant cleave DNA in a single strand of the recognition site. Also, \textit{Eco}RV mutants are not affected in ground state binding but rather in the stabilization of the transition state, and catalysis is significantly altered compared to binding when the symmetry of the protein-DNA interface is disturbed. Taken together, these data suggest that an asymmetry in the enzyme is manifested in the catalytic centers of the two subunits only in the transition state, and that a nonlocal pathway---in which some physical quantity is conserved---may be used for coordination. Under physiologically optimum conditions, several type II endonucleases demonstrate products that are cleaved entirely in both strands without producing intermediate single-strand cuts \cite{Pingoud3, Pingoud2, Halford, Maxwell}, suggesting a mechanism of synchronization between spatially separated nucleotides that is conserved in this class of enzymes. Such an absolute correlation over distance is the hallmark of quantum entanglement \cite{EPR}.

The specific biochemical approaches employed by these enzymes to attack the phosphodiester bonds of the DNA helix are studied extensively in the literature. Here, we seek to describe a simple yet plausible model for the fundamental physical phenomena underlying the process of ``substrate-assisted'' catalytic synchronization. We propose that a subset of type II restriction endonucleases coordinate their two catalytic centers by entangling electrons in the target phosphodiester bonds through dipole-dipole couplings in the bound DNA sequence. These bonds are the prime candidates for the location of the entangled electrons, as the close proximity of at least one Mg$^{2+}$ ion to each bond \cite{Horton} delocalizes the charge distribution and stabilizes the transition state.

\begin{figure}[htb]
\centering
\includegraphics[width=0.9\textwidth]{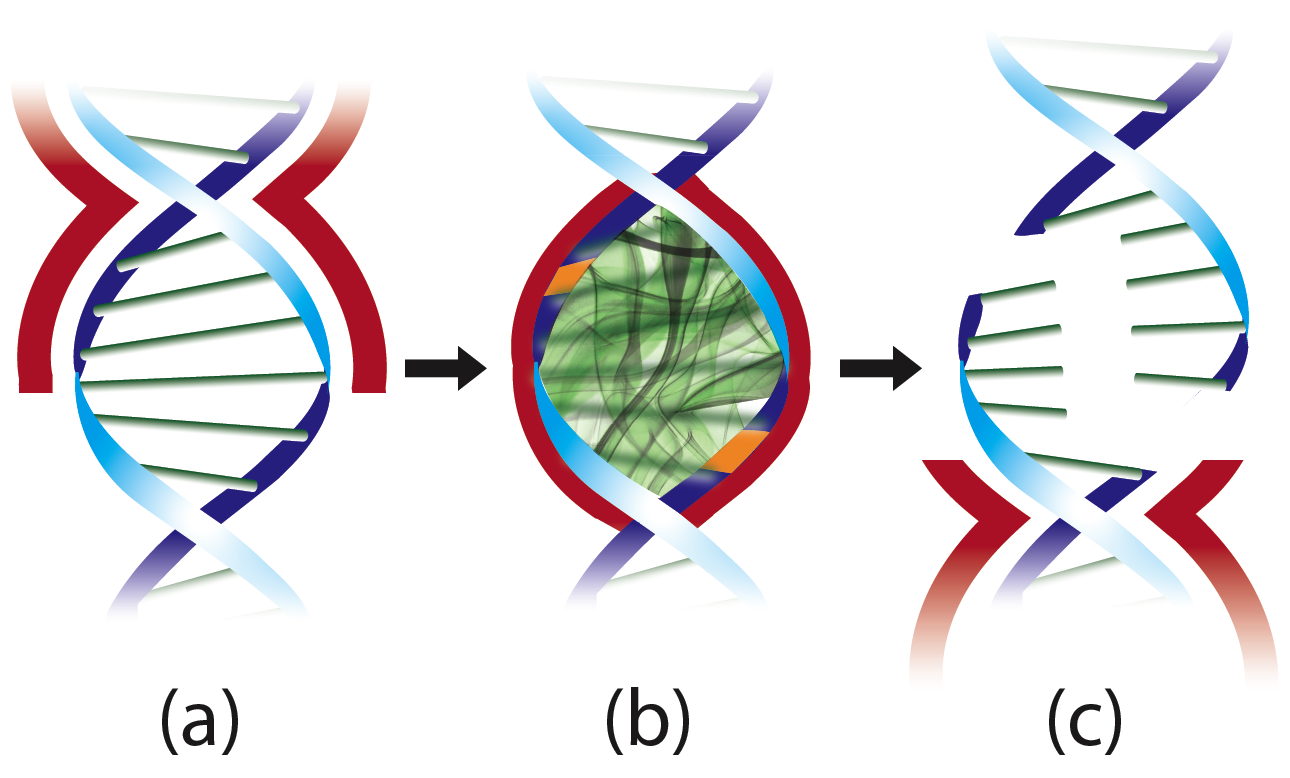}
\caption{\textbf{Proposed quantum entanglement in orthodox type II restriction endonuclease catalysis.}          
(a) Enzyme (red) searches for recognition sequence on DNA (blue) by facilitated diffusion. (b) Enzyme recognizes target site, undergoing conformational change to tightly bind the DNA sequence and form the decoherence-free subspace. Clamping induces excitation of quantized oscillations from coupled base-pair electron clouds (green), entangling two electrons in phosphodiester bonds (orange) on opposing strands of the helix. (c) Synchronized catalysis occurs as quanta decay symmetrically into the entangled bonds, thus breaking the DNA helix in a single binding event. Dissociation of the complex produces characteristic ``sticky ends'' with single bases exposed.}
\label{fig:Quantum_helix}
\end{figure}

We must address the question posed in the opening paragraph, regarding biology's ability to exploit macroscopic quantum effects. Type II restriction endonucleases that cut DNA in a concerted manner, as shown in Figure \ref{fig:Quantum_helix}, would maintain quantum coherence in the DNA substrate by acting as a decoherence shield upon specific binding. Decoherence shields have been evoked in the quantum biology literature, particularly in photosynthetic complexes, using both quantitative \cite{chin2013role} and qualitative \cite{whaley2011quantum} arguments. We base our proposal on the experimentally confirmed observation that the conformational change induced in these enzymes between nonspecific and specific binding is commensurate with the exclusion of ions and over 100 water molecules from the surface of DNA \cite{Water1, Water2, Water3}. Water, the necessity of life, is the primary vehicle by which thermal agitation is transmitted to biological systems, and squeezing water away from the DNA helix may be the procedure by which orthodox type II restriction endonucleases create a decoherence-free subspace for quantum entanglement to occur. The inverse-squared nature of Coulomb's law ensures that every doubling of the distance from charge reduces the electrostatic force by a factor of four. Orthodox type II endonucleases would exploit this feature when forming decoherence shields around DNA, extending the separation between the helix core and counterions. Charged amino acid residues on the surface of the enzymes stabilize the ions, further neutralizing the local electromagnetic environment. Release of charge-countering ions from the protein-DNA interface and charge cancellation by amino acid residues would ensure that electromagnetic interaction with the delicately shielded quantum state would be kept to a minimum. 

\begin{figure}[htb]
  \begin{center}
  \subfigure[\,Debye length in water]{
            \label{80permit}
            \includegraphics[width=0.63\textwidth]{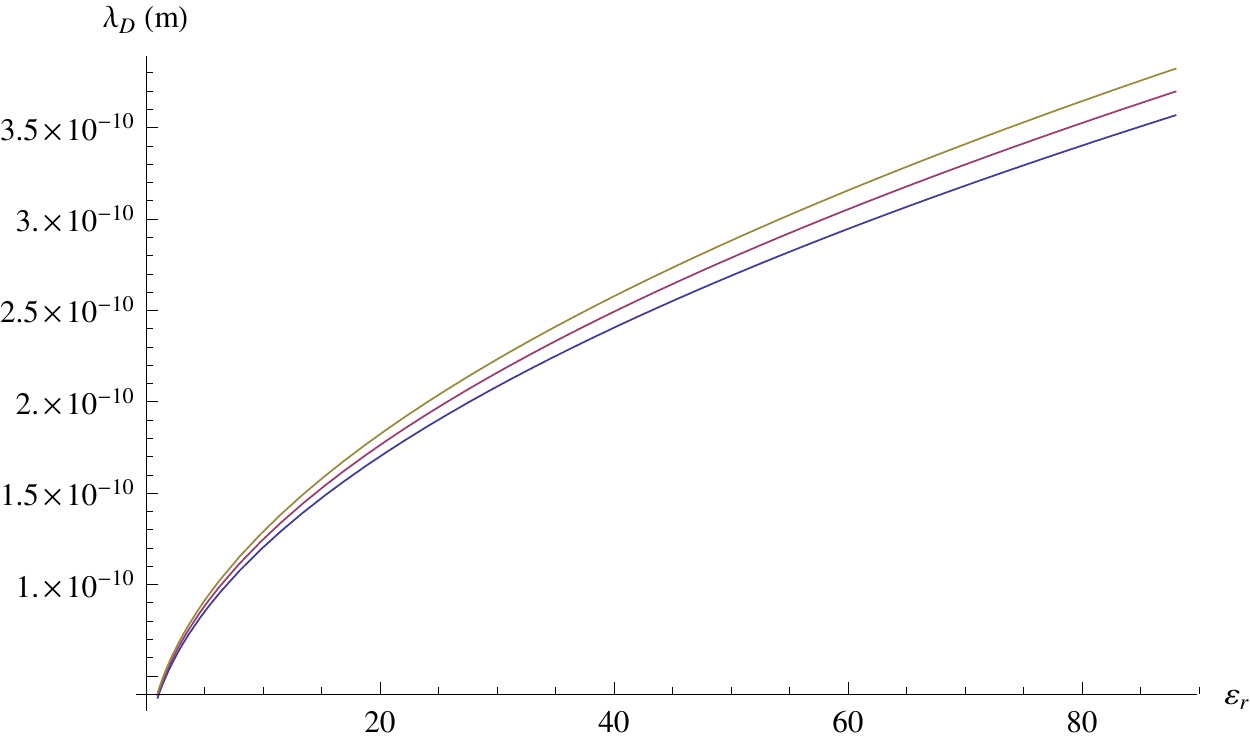}}
   \subfigure[\,Debye length in quasi-vacuum]{
           \label{20permit}
          \includegraphics[width=0.63\textwidth]{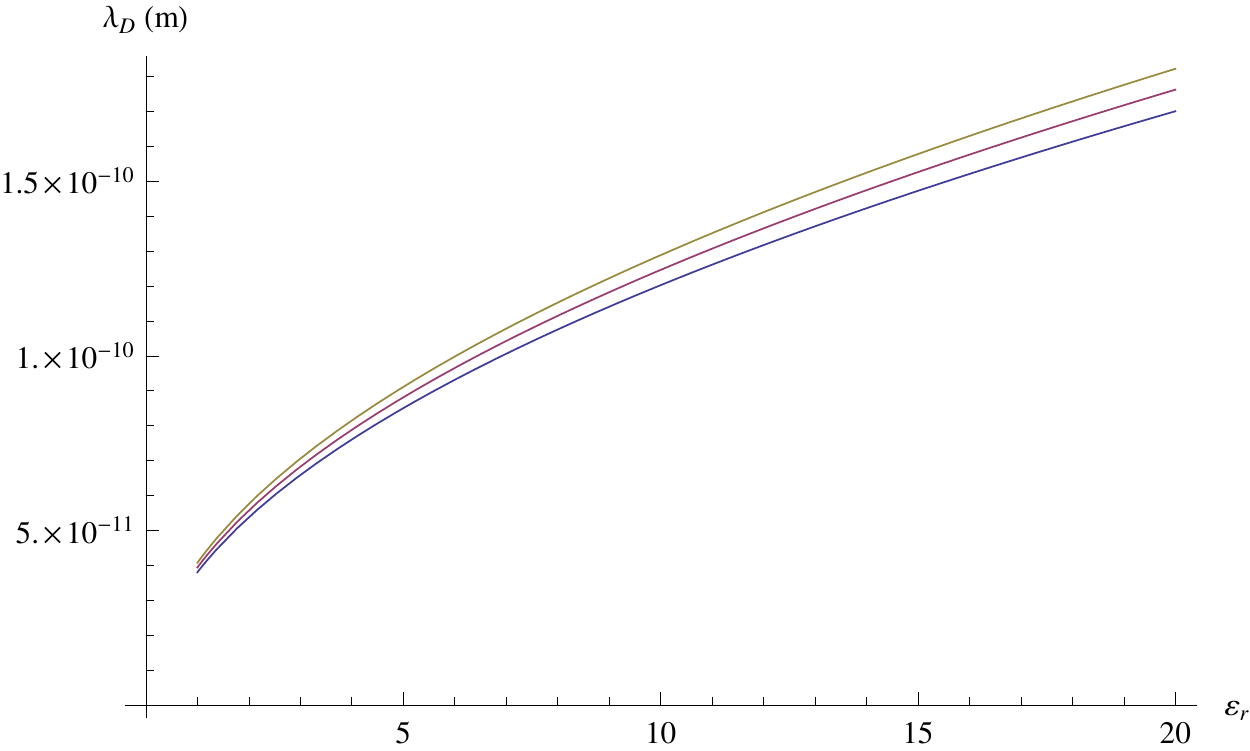}}
  \end{center}
    \caption{The Debye screening length (Equation \ref{Debyelength}) as a function of relative permittivity across the full physiological domain from vacuum to water at freezing (a) and across a limited domain closer to the ideal vacuum limit (b). Curves are shown at $T=270\,$K (blue), $T=290\,$K (red), and $T=310\,$K (gold). Notice that $\lambda_D$ varies with $\sqrt{\varepsilon_r}$, and that in quasi-vacuum this length approaches the order of half an angstrom, or about $1/7$ of the distance between base pairs. For comparison, the Debye length for physiological concentrations of MgCl$_2$---including the chlorine ions---in water at $T=300\,$K is $1.76\,$nm, a little more than the length of five DNA base pairs.}
   \label{Debyegraphs}
\end{figure}

The Debye screening length (also known as the Debye-H\"{u}ckel length) is a good measure for determining the characteristic scale far outside which electrostatic effects are exponentially screened. It is given in units of meters by
\begin{equation} \label{Debyelength}
\lambda_D=\left(8\pi \ell_B N_A I\right)^{-\frac{1}{2}},
\end{equation}
where $\ell_B=q_e^2/(4\pi\varepsilon k_B T)$ is the Bjerrum length, $N_A = 6.022 \, 14\,... \times 10^{23} \, \text{mol}^{-1}$ is the Avogadro constant, and $I=\frac{1}{2}\sum \limits_{i=1}^n c_i z_i^2$ is the ionic strength, with $c_i$ as the concentration $\left(\frac{\text{mol}}{\text{m}^3}\right)$ and $z_i=q_i/q_e$ as the integer charge number of the $i$th ion species in solution. Here $\varepsilon$, the absolute permittivity in media, is the product of the relative permittivity $\varepsilon_r$, a dimensionless dielectric constant, and the permittivity of free space $\varepsilon_0=8.854\,187\, 817\, 620... \times 10^{-12}$ farads per meter. Taking the diameter of DNA as $23.7\,$\r{A} and the length of a six-bp sequence as $20.4\,$\r{A}, we can compute the concentration in a notional cylindrical volume for the Mg$^{2+}$ ions positioned after the restriction endonuclease has squeezed water and other counterions away from the DNA surface. The results for the Debye length of this model system are presented in Figure \ref{Debyegraphs} and suggest that these ions exert an enormous screening effect beyond even a small fraction of the inter-base-pair distance in the local quasi-vacuum.

Researchers have investigated vibrational modes in DNA from theoretical \cite{Cocco}, computational \cite{Blinov}, and experimental \cite{Woolard} perspectives. In the following sections, we describe our physical model for DNA sequences that unlike these past investigations exploits their electronic quantum vibrational modes, which impact the chemistry. We then demonstrate the agreement between the predictions of our mathematical model and existing data on the energetics of type II restriction endonuclease catalysis. Finally, we conclude with thoughts on the applicability of our model to palindromic inverted repeat sequences in the genome.

\section{Physical Model of DNA Sequence}
Similarly to (yet distinct from) previous authors \cite{Rieper}, we simulate collective electronic behaviors within the DNA helix through the interactions between delocalized electrons. These electrons belong to the planar-stacked base pairs that serve as ``ladder rungs'' stepping up the longitudinal helix axis. Coulombic interactions between electron clouds proceed by induced dipole formation due to London dispersion (of van der Waals type) from nearest neighbors. Each rung of the double-stranded helix---composed of either an A:T or C:G base pair---can thus be visualized as an electronically mobile sleeve vibrating with small perturbations around its fixed positively charged core, and the DNA helix thus becomes a chain of electronically coupled harmonic oscillators with twist angle $\theta$ radians per base pair. 

The Hamiltonian for such a molecule of length $N$ nucleotides is
\begin{equation} \label{eq:Hamiltonian}
H=\sum^{N-1}_{s=0}\frac{\mathbf{p}_s^2}{2m_s} + \frac{m_s}{2}\left(\omega_{s,xx\,}^{2}x_s^2 +\omega_{s,yy\,}^2y_s^2+\omega_{s,zz\,}^2z_s^2\right ) + V_s^{int},
\end{equation}
where $\mathbf{r}_s=(x_s, y_s, z_s)$ are the displacement coordinates between each electron cloud and its base-pair core, the coordinates $x_s, y_s$ span the transverse plane of each base-pair cloud, and the $z_s$ are aligned along the longitudinal axis. The dipole-dipole interaction terms are given by
\begin{equation} \label{eq:dip-dip}
V_s^{int} = \frac{1}{4\pi\epsilon_0 d^3}\left[\gv{\pi}_s \cdot \gv{\pi}_{s+1}- \frac{3(\gv{\pi}_s \cdot \mathbf{d})(\gv{\pi}_{s+1} \cdot \mathbf{d})}{d^2}\right],
\end{equation}
where $\mathbf{d} = d\hat{\mathbf{z}}$ connects the centers of nearest-neighbor base-pair dipoles $\gv{\pi}_s = Q\mathbf{r}_s$ and $\gv{\pi}_{s+1} = Q\mathbf{r}_{s+1}$.

\begin{table}
\centering
\begin{tabular}{ |c || c | c | c| } 
\hline
bp & $\omega_{xx}$ & $\omega_{yy}$ & $\omega_{zz}$ \\ \hline
A:T & 3.062 & 2.822 & 4.242\\ \hline
C:G & 3.027 & 2.722 & 4.244\\ \hline
\end{tabular}
\caption{DNA base pair electronic angular frequencies, in units of $10^{15}$ radians per second.}
\label{DNAbpomega}
\end{table}

As shown in Table \ref{DNAbpomega}, $\omega_{s,ii\,}$ are the diagonal elements of the angular frequency tensor for each base-pair electronic oscillator and are determined from electric polarizabilities ($\alpha_{A,\,ii},\alpha_{C,\,ii}, \alpha_{G,\,ii}, \alpha_{T,\,ii}$) of the individual nucleotide bases:
\begin{equation} \label{eq:omegatensor}
\omega_{A:T, \,ii} = \sqrt{\frac{Q^2}{m(\alpha_{A, \,ii}+\alpha_{T,\, ii})}}, 
\end{equation}
and similarly for $\omega_{C:G, \,ii}$. Equation \ref{eq:omegatensor} may be derived from the fundamental dipole relation $\gv{\pi}=\gv{\alpha}\cdot \mathbf{E}$, where the dipole moment is proportional to the electric field by the diagonal rank-two polarizability tensor $\gv{\alpha}$ unique for each base, and by equating the electric Lorentz and spring forces. The numerical elements for each $\gv{\alpha}$ were of course used in constructing Table  \ref{DNAbpomega}. Mathematically, using matrix elements for the derivation,
\begin{eqnarray}
\nonumber &\gv{\pi} =\gv{\alpha}\cdot \mathbf{E} = \gv{\alpha}\cdot \frac{\mathbf{F}}{Q} \\
\nonumber &\downarrow\\
\nonumber &Qr_t = \alpha_{tu}\frac{k_{uv}r_v}{Q} \\ 
\nonumber &\downarrow\\
\nonumber &Q^2 \delta_{tv} = \alpha_{tu}k_{uv} \\ 
\nonumber &\downarrow\\
&Q^2\gv{\alpha}^{-1} =\mathbf{k} = m\gv{\omega}^2,
\end{eqnarray}
which yields precisely the matrix version of Equation \ref{eq:omegatensor}. The tensor elements $\alpha_{ii}$ account for anisotropies, which have been determined from perturbation theory \cite{Fock-Dirac}, simulation \cite{CNDO}, and experiment \cite{Basch} generally to within five percent agreement.  The mass and charge of an electron are used because, in the spatially separated phosphodiester bonds, single electrons would be entangled through the intermediary DNA couplings. In other words, an electron in the sugar-phosphate spiral is entangled with a distant electron via nearest-neighbor couplings through the core of the DNA helix.

\begin{figure}[htb]
\centering
\includegraphics[width=1.0\textwidth]{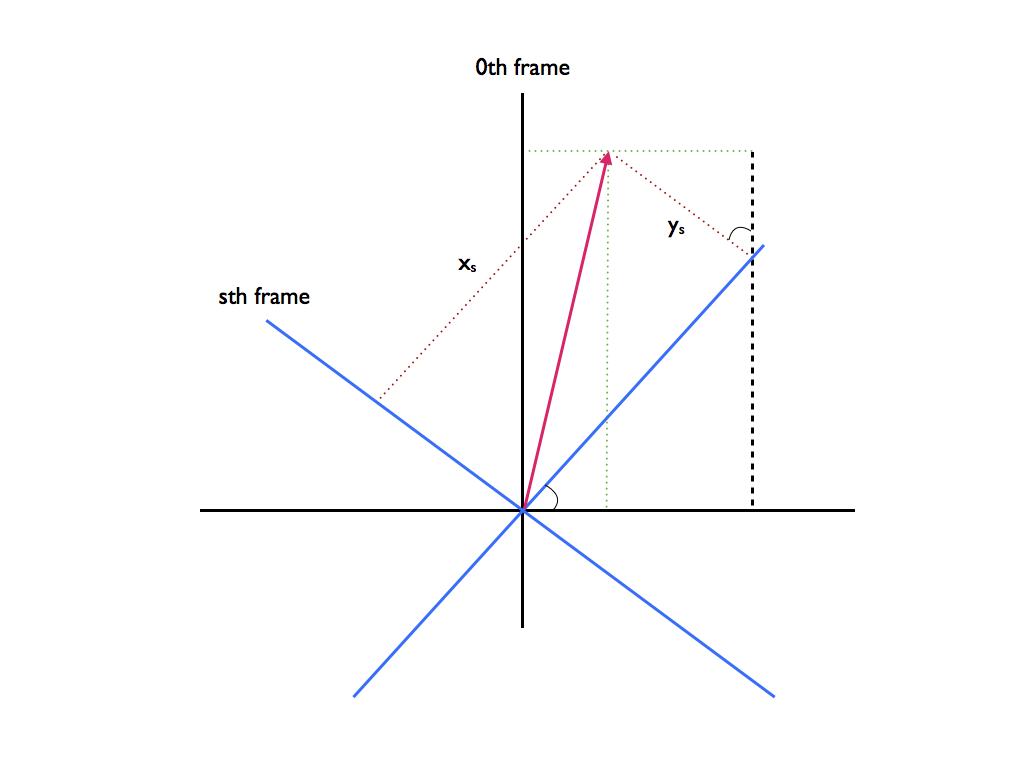}
\caption{Rotated frames for a sequence of DNA base pairs. The reference frame of the $s$th base pair (blue, solid) is rotated by $s\theta$ from the $0$th base-pair reference frame (black, solid) due to the twist in the DNA helix. Identifying the $s$-frame components (orange, dotted) of the displacement vector (red, arrow) as $(x_s, y_s)$, we can derive the $0$-frame components (green, dotted, and black, dashed) from the trigonometry of the formed triangles.}
\label{fig:bpAxes}
\end{figure}

Due to the twist in the helix about the longitudinal axis, we must account for cross terms between directional components of two interacting dipoles. Choosing a single coordinate frame $(X, Y, Z)$ that corresponds with $(x_0, y_0, z_0)$ of the $0$th base pair, we now write the interaction potential for the $s$th electronic oscillator as
\begin{equation} \label{eq:singleframe}
V_s^{int} = \frac{Q^2}{4\pi\epsilon_0 d^3}\left[X(s)X(s+1)+Y(s)Y(s+1)-2Z(s)Z(s+1) \right].
\end{equation}
We can return to the original displacement components of Equation \ref{eq:Hamiltonian} by the transformation shown in Figure \ref{fig:bpAxes}:
\begin{align}
\nonumber X(s) &= x_s \cos s\theta - y_s \sin s\theta\\ \nonumber
Y(s) &= x_s \sin s\theta + y_s \cos s\theta\\
Z(s) &= z_s,
\end{align}
substituting into Equation \ref{eq:singleframe} to obtain 
\begin{equation} \label{eq:dipolepotential}
V_s^{int} = \frac{Q^2}{4\pi\epsilon_0d^3}[x_s x_{s+1} \cos \theta + y_s y_{s+1} \cos \theta + (y_s x_{s+1} - x_s y_{s+1}) \sin \theta- 2z_s z_{s+1}],
\end{equation}
where the orientation of the helix and the twist angle are reflected in the different factors for the quadratic couplings. 

Any physical model that seeks to simplify biological complexity to a finite set of parameters will make certain assumptions. The most prominent simplification we make is treating the DNA after specific binding as separate from the enzyme and computing normal modes from the electronic properties of the bases alone. Furthermore, the general $xyz$ coupling consists of additional dipole cross terms arising from the DNA kink angle relative to the longitudinal axis. This coupling requires diagonalizing a $3N \times 3N$ matrix, increasing the number of determinant operations by a factor of $\sim(3N)!/(2N)!$, or nearly $13.5$ million times for six-bp recognition sequences. 

In the case of an infinite helix composed of homogeneous base pairs, we may transform the infinite sums into easily calculable integrals. From the idealized Hamiltonian
\begin{equation}
H_I= \frac{1}{2M} \left \{ \sum \limits_{n=-\infty}^{+\infty} p_n^2 \right \} + \frac{1}{2}M \omega^2 \left \{ \sum \limits_{n=-\infty}^{+\infty} q_n^2 \right \} +\frac{1}{2}\Gamma \left \{ \sum \limits_{n=-\infty}^{+\infty} (q_n-q_{n+1})^2 \right \},
\end{equation}
where the uniform $M$ and $\omega$ reflect the chain homogeneity and $p_n(t), q_n(t)$ are the deviation from equilibrium for the $n$th component oscillator in a single dimension with arbitrary interaction potential $\Gamma$, we may use the Bessel-Parseval relation \cite{Cohen-Tannoudji} and recursion to obtain the diagonalized form
\begin{equation} \label{eq:normalmodes}
H_I=\frac{\ell}{2\pi} \int \limits_{-\frac{\pi}{\ell}}^{+\frac{\pi}{\ell}} dk \left \{\frac{1}{2M}|P_0(k, t)|^2 + \frac{1}{2} M \left [\omega^2 + \frac{4\Gamma}{M} \sin^2\left ( \frac{k \ell}{2} \right) \right] |Q_0(k, t)|^2 \right\},\end{equation}
where $\ell$ is the unit distance of the oscillator chain and the new position and momenta coordinates are defined using Fourier series:
\begin{eqnarray} \label{newp&q}
\begin{gathered}
Q_0 = \sum \limits_{n=-\infty}^{+\infty} q_n e^{-ink \ell},\\
P_0 = \sum \limits_{n=-\infty}^{+\infty} p_n e^{-ink \ell},\\
\end{gathered}
\hspace{0.25em}
\begin{gathered}
Q_1 = \sum \limits_{n=-\infty}^{+\infty} q_{n+1} e^{-ink \ell},...\\
P_1 = \sum \limits_{n=-\infty}^{+\infty} p_{n+1} e^{-ink \ell},...\\
\end{gathered}.
\end{eqnarray}

The normal modes of vibration are obtained readily from Equation \ref{eq:normalmodes}: $\Omega(k)=\sqrt{\omega^2 + \frac{4\Gamma}{M} \sin^2\left ( \frac{k\ell}{2} \right)}$, which applies for any infinite coupled oscillator sequence with homogeneous components in the first Brillouin zone, where the wave vector obeys $-\frac{\pi}{\ell} \leq k \leq \frac{\pi}{\ell}$. Transforming coordinates by Fourier expansion thus allows us to convert the idealized Hamiltonian, which was expressed as an infinite sum of coupled oscillators, into a definite integral over uncoupled collective modes for each physically distinguishable state. 

Though a helpful comparison, such an analytical solution is not rigorously applicable for finite, real DNA sequences. It does not show the complement of modes depressed below the homogeneous trapping frequency $\omega$ because the generality of the real-valued function $\Omega$ would have to be restricted for specified values of $\Gamma$. However, numerical analysis will certainly suffice for genomic length scales. For recognition sequences of four to eight base pairs, where type II restriction endonucleases tightly enforce boundary conditions, little must be done in the way of implementing numerical recipes.

\section{Energetics of Catalysis}
Our computations for type II recognition sequences begin with standard finite matrix methods. Separating Equation \ref{eq:Hamiltonian} into energy contributions from transverse ($H_{xy}$) and longitudinal ($H_z$) modes, we may write the symmetric longitudinal potential matrix for a four-bp sequence as
\begin{equation}
\mathbf{V}_z =
{\begin{pmatrix}
k_{1,zz} & \gamma_{12}^z & 0 & 0 \\
\gamma_{21}^z  & k_{2,zz} & \gamma_{23}^z  & 0 \\ 
0 & \gamma_{32}^z  & k_{3,zz} & \gamma_{34}^z  \\
0 & 0 & \gamma_{43}^z  & k_{4,zz} \\
\end{pmatrix}},
\end{equation}
where $k_{1,zz}=m_1\omega_{1,zz}^2$, etc., and $\gamma_{s,s+1}^z=\gamma_{s+1,s}^z=-\frac{Q^2}{2\pi\epsilon_0 d^3}$ denotes the $z_s z_{s+1}$ coefficient from Equation \ref{eq:dipolepotential}. The symmetric transverse potential matrix is
\begin{equation}
\mathbf{V}_{xy} =
{\begin{pmatrix}
k_{1,xx}& \gamma_{12}^x & 0 & 0 &0 & \gamma_{12}^{xy} &0 &0\\
\gamma_{21}^x  & k_{2,xx} & \gamma_{23}^x  & 0  & \gamma_{21}^{xy} &0 &\gamma_{23}^{xy}  &0\\ 
0 & \gamma_{32}^x  & k_{3,xx} & \gamma_{34}^x &0 &  \gamma_{32}^{xy}  &0 & \gamma_{34}^{xy}\\
0 & 0 & \gamma_{43}^x  & k_{4,xx} &0 & 0& \gamma_{43}^{xy} & 0 \\
0&\gamma_{12}^{yx}&0&0&k_{1,yy}&\gamma_{12}^y&0&0\\
\gamma_{21}^{yx}&0&\gamma_{23}^{yx}&0&\gamma_{21}^y&k_{2,yy}&\gamma_{23}^y&0\\
0&\gamma_{32}^{yx} &0&\gamma_{34}^{yx} &0&\gamma_{32}^y&k_{3,yy}&\gamma_{34}^y\\
0&0&\gamma_{43}^{yx} &0&0&0&\gamma_{43}^y&k_{4,yy}\\
\end{pmatrix}},
\end{equation}
where $k_{1,xx}=m_1\omega_{1,xx}^2$, etc., and $\gamma_{12}^{xy} =-\frac{Q^2 \sin\theta}{4\pi\epsilon_0 d^3}=-\gamma_{21}^{xy}$, etc. The diagonal kinetic matrix $\mathbf{T}$ consists of the electronic oscillator masses:
\begin{equation}
\mathbf{T}_z =
\text{diag}(m_1, m_2, m_3, m_4), \quad
\mathbf{T}_{xy}=
\text{diag}(m_1, m_2, m_3, m_4, m_1, m_2, m_3, m_4). 
\end{equation}
In type II endonuclease recognition sequences, $m_s = m_e$ for all $s$. The problem then consists of solving
\begin{equation} 
\det(\mathbf{V}_j - \Omega_{s,j}^2\mathbf{T}_j) = 0
\label{eq:goldstein}
\end{equation}
for the eigenfrequencies $\Omega_{s,j}$, which are in general complex-valued. The closed-form solutions are rather cumbersome unless homogeneity is assumed in the sequence $(m_s=m, \omega_{s,zz}=\omega, \gamma_{s,s+1}^z = \gamma)$. The longitudinal mode frequencies in the four-bp case simplify to
\begin{align} \label{eq:goldenrat}
\nonumber \Omega_{0,z}^{\,2} &= \omega^2-\varphi \frac{\gamma}{m}\\
\nonumber \Omega_{1,z}^{\,2} &= \omega ^2- (\varphi-1)\frac{\gamma }{m}\\
\nonumber \Omega_{2,z}^{\,2} &= \omega ^2+ (\varphi-1)\frac{\gamma }{m}\\
\Omega_{3,z}^{\,2} &= \omega ^2+ \varphi \frac{\gamma}{m},
\end{align}
where $\varphi = \frac{1+\sqrt{5}}{2}$ is the golden ratio. 

Complex-valued roots are taken to be decaying states, with the real part giving rise to the reported value of $\Omega_{s,j}$. The time scales for decay of these imaginary frequencies are $O(10^{-16})$ or $O(10^{-15})$ seconds. However, these complex roots generally exhibit strictly an imaginary contribution---with no real part---and can therefore be neglected. The real-valued frequencies with zero decay constant dominate on the time scale of the oscillatory synchronization.

It can be shown that these classically derived normal-mode frequencies correspond to the normal-mode frequencies in the quantum case \cite{BriggsEisfeld}, as both diagonalization procedures correspond to the same combination of coordinate rescaling and $SO(N)$ rotation. We introduce the normal-mode lowering operator 
\begin{equation}
a_{s, j}=\sqrt{\frac{m\Omega_{s,j}}{2\hbar}}(\v{r}^\prime_s)_j+ \frac{i}{\sqrt{2m \hbar \Omega_{s,j}}}(\v{p}^\prime_s)_j,
\end{equation} 
where the primed coordinates are formed---similarly to Equation \ref{newp&q}---from a finite linear combination of the original displacements and momenta:
\begin{equation}
(\v{r}^\prime_s)_j = \sum \limits_{n=0}^{N-1} (\v{r}_n)_j \, \exp\left(-\frac{2\pi i ns}{N}\right), \, \, \, \, \, (\v{p}^\prime_s)_j = \sum \limits_{n=0}^{N-1} (\v{p}_n)_j \, \exp \left(-\frac{2\pi i ns}{N}\right). 
\end{equation}
Our Hamiltonian thus takes the standard diagonalized form
\begin{equation}
H_j= \sum^{N-1}_{s=0}\hbar\Omega_{s,j} \left (a_{s,j}^\dagger a_{s,j} +\frac{1}{2} \right),
\end{equation}
where $a_{s,j}^\dagger$ is the raising operator for the $s$th normal mode of the collective electronic oscillations for the $j=xy$ or $j=z$ potential. The eigenstates of $H_j$ are given by
\begin{equation}
\ket{\psi_{s,j}}=a_{s, j}^\dagger \ket{0},
\end{equation}
where $s=0,1,\dots,N-1$ for the $j=xy, z$ potential. We will examine only the lowest energy states because these modes are the most easily excited.

\begin{table}
\centering
\begin{tabular}{ |c||c|c|c|c| }
\hline
ÊÊÊÊÊÊÊÊ$d$                    ÊÊ & $\hbar\Omega_{0,xy}/2$ & $\hbar\Omega_{1,xy}/2$ & $\hbar\Omega_{0,z}/2$ & $\hbar\Omega_{1,z}/2$Ê \\ \hline
        $3.0\,$\r{A} &   2.86       &  3.07     &   4.45  & 7.35 \\ \hline
      Ê$3.2\,$\r{A} &ÊÊÊ3.11ÊÊÊÊÊÊÊ&ÊÊ3.31ÊÊÊÊ &   4.78  & 7.14 \\  \hline
        $3.4\,$\r{A} &ÊÊÊ0.53ÊÊÊÊÊÊÊ&ÊÊ1.35ÊÊÊÊ &   1.99  & 5.02 \\  \hline
        $3.6\,$\r{A} &ÊÊ1.70ÊÊÊÊÊÊ &ÊÊ2.11ÊÊÊÊ &   3.02  & 5.20 \\  \hline
        $3.8\,$\r{A} &ÊÊÊ0.40ÊÊÊÊÊÊÊ&ÊÊ0.82ÊÊÊÊÊÊ&  1.67  & 3.65 \\  \hline
\end{tabular}
\caption{\textit{Eco}RI DNA recognition sequence (GAATTC) zero-point modes, in units of $\varepsilon_{P-O} \simeq 0.23$ eV, as a function of inter-base-pair spacing $d$. The helix twist angle $\theta \simeq \pi/5$ is constant for the six-bp sequence.} 
\label{tab:EcoRI}
\end{table}

The energy required \textit{in vivo} to break a single phosphodiester bond is $\varepsilon_{P-O} \simeq 0.23$ eV \cite{Phospho}, which is less than two percent of the energy required to ionize the hydrogen atom but about ten times the physiological thermal energy ($k_B T$). This suitable value of $\varepsilon_{P-O}$ is comparable with the quantum of biological energy released during nucleotide triphosphate hydrolysis; it ensures that the bonds of the DNA backbone are not so tight as to be unmodifiable but remain strong enough to resist thermal degradation. Remarkably, for \textit{Eco}RI the difference in free energy between the nonspecific and specific complex (i.e., clamping energy) is approximately $2\varepsilon_{P-O}$. As shown in Table \ref{tab:EcoRI}, several zero-point modes for GAATTC are resonant with this clamping energy, and at standard inter-base-pair spacing of $3.4\,$\r{A} we calculate the ground state $z$ mode to within $0.5\%$ of the energy required for double-strand breakage. Enzyme clamping thus imparts the quanta of energy needed. The clamping effect on the helix would excite the lowest-energy $z$ mode in the DNA sequence, which transports the quanta along the longitudinal axis to break two phosphodiester bonds simultaneously.

\begin{figure}[htb]
  \begin{center}
  \subfigure[\,dsAAAA longitudinal modes]{
            \label{fig:AAAAz}
            \includegraphics[width=0.63\textwidth]{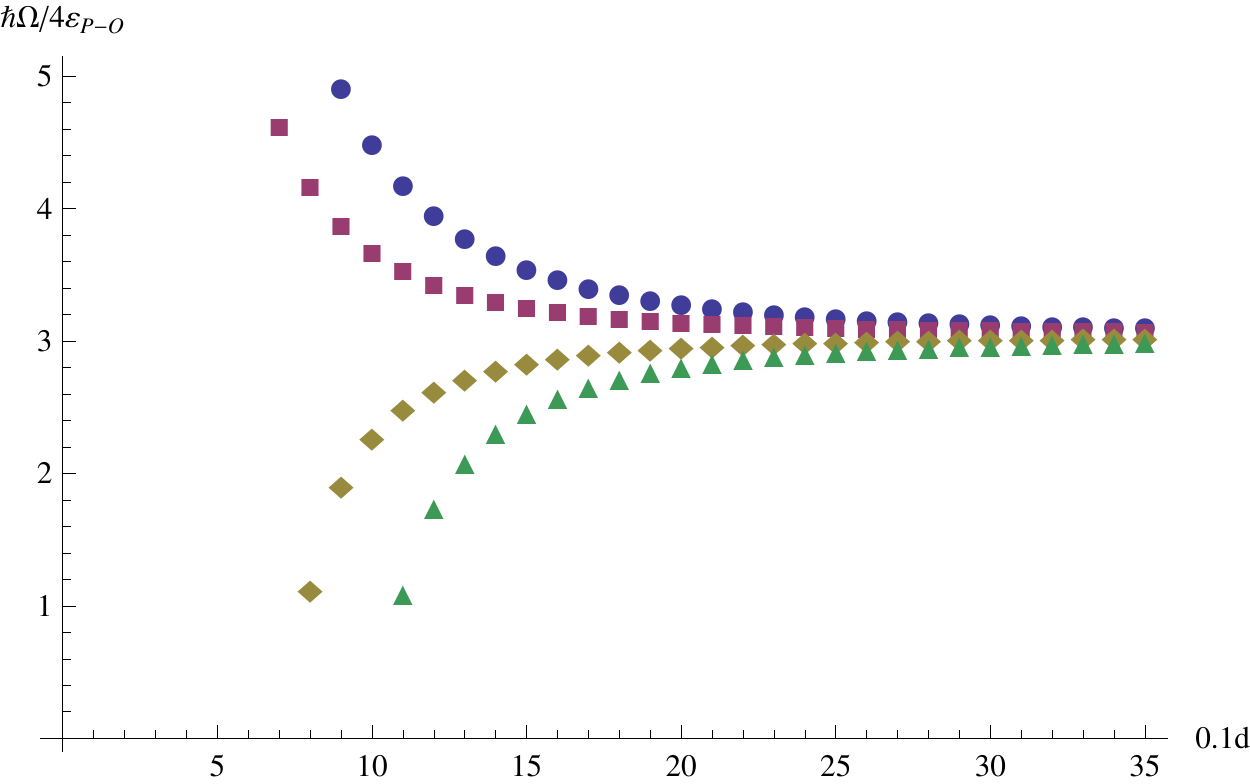}
        }
        \subfigure[\,dsAAAA transverse modes]{
           \label{fig:AAAAxy}
          \includegraphics[width=0.63\textwidth]{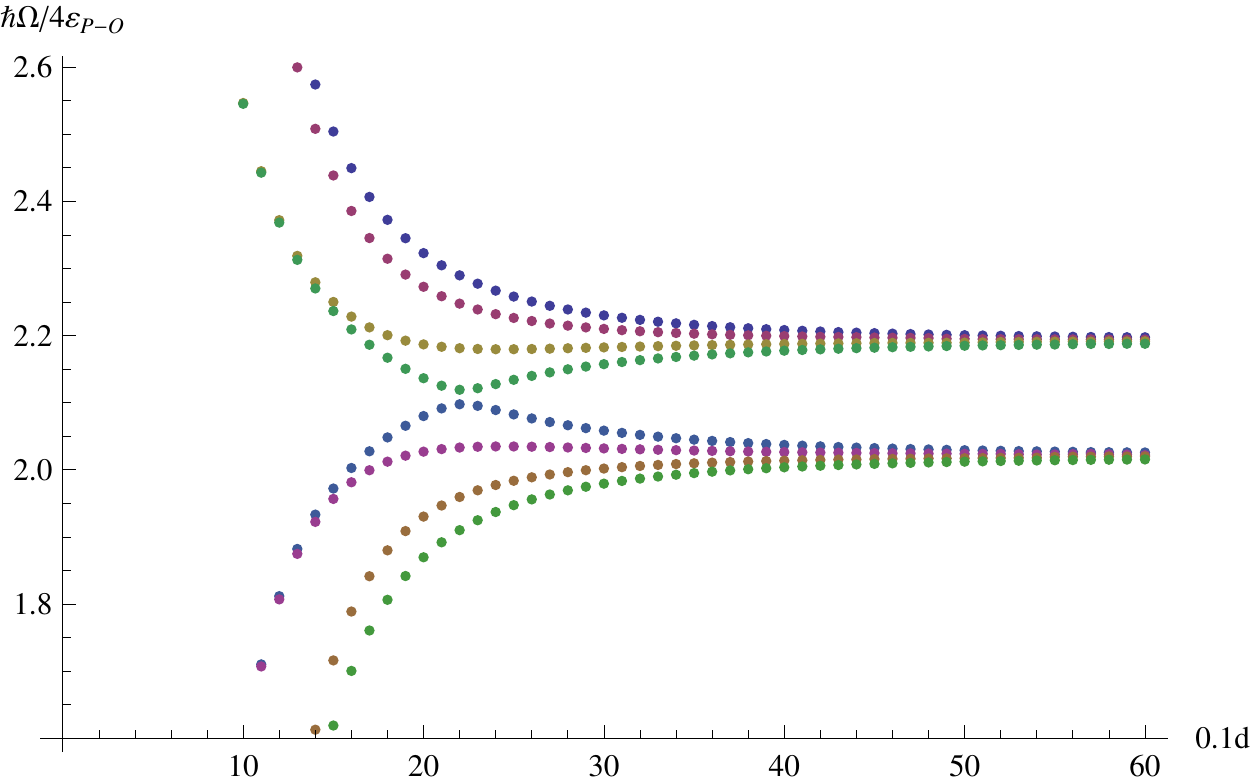}
        }
  \end{center}
    \caption{Longitudinal (a) and transverse (b) normal mode frequencies parametrized by the inter-base-pair spacing for the double-stranded sequence AAAA. Notice the appearance of the ground state normal modes (green triangles in (a), light green circles in (b)) at approximately the observed equilibrium base-pair spacing $d$, where for more compacted DNA sequences these modes remain complex-valued. The longitudinal ground state mode at $d$, with $\hbar\Omega/2 \approx 2.0 \varepsilon_{P-O}$, is suitably equipped for the task of DNA double-strand breakage. $\varepsilon_{P-O} \simeq 0.23$ eV, $d=3.4\,$\r{A}.}
   \label{subfig:dsAAAA}
\end{figure}

Parametrization of our model by the inter-base-pair spacing is shown in Figure \ref{subfig:dsAAAA} for a homogeneous dsDNA sequence of four base pairs. What becomes quickly apparent in Figure \ref{fig:AAAAz} is the rapid convergence of the longitudinal modes to $\hbar\Omega/2 \approx 6.0 \varepsilon_{P-O}$, within less than three times the standard observed inter-base-pair spacing, suggesting that this parameter has been evolutionarily optimized to maximize variation in DNA electronic oscillations at physiologically relevant length scales. Similar behavior is observed for the transverse modes in Figure \ref{fig:AAAAxy}, with a bi-modal convergence around $\hbar\Omega/2 \approx 4.4 \varepsilon_{P-O}$ and $4.0 \varepsilon_{P-O}$. Complex-valued frequencies describe ephemeral, decaying states not on par with the more permanent, real-valued modes.

Surprisingly, the modes for AAAA in Figure \ref{subfig:dsAAAA} converge to the same values as the triplet codon case (dsAAA, data not shown) for the zero-point energies over comparable length scales. The middle harmonics bifurcate quickly and diverge when the spacing dips below $5.0$\r{A} (Figure \ref{fig:AAAAz}) and $7.0$\r{A} (Figure \ref{fig:AAAAxy}). What these data suggest is that, if DNA were ``tuned'' away from this equilibrium spacing, then we would not observe the plethora of distinct coherent vibrations in DNA nor would certain collective modes achieve energies of significance to genomic and biological metabolism. 

This possibility of energy transfer from the tight binding of the recognition sequence to the catalytic transition state has been raised by previous authors \cite{Jen-Jacobson}. The oscillation mode of the bound DNA entangles electrons in the two phosphodiester bonds, as illustrated schematically in Figure \ref{fig:Quantum_helix}. One could think of it as the phonon-like quanta mediating the entanglement between electrons. Breaking these bonds or dissociating the complex with no strand breaks is akin to state measurement, when quantum coherence is lost between the identical observables. The enzyme's ``measurement'' of quantum state outcomes $\ket{cc},\ket{cn},\ket{nc},\ket{nn}$---where $\ket{c}$ corresponds to a single-strand catalytic event and $\ket{n}$ to no catalytic event---bears witness to the total production of cut-symmetric states $\ket{cc}, \ket{nn}$ to the exclusion of cut-asymmetric states $\ket{cn},\ket{nc}$ under optimal conditions. (This heuristically defined basis can be rewritten as a complex linear combination of spherical harmonics, owing to the hybridized electron orbitals involved.) Thus the entangled two-qubit Bell state $\alpha\ket{cc} \pm \beta\ket{nn}$ for some $\alpha\neq 0, \, \beta\neq 0, \, \abs{\alpha}^2+\abs{\beta}^2=1$ is maintained in the decoherence-free subspace until double-strand breakage or dissociation of the restriction endonuclease from the DNA occurs. The fundamentally quantum zero-point mode provides the energy that would otherwise have been requisitioned from a high-energy molecule such as ATP or AdoMet. 

The entangled state, represented above in the ``readout'' basis, can be equivalently described in the energy basis. Writing the $0$th transverse mode and the $1$st longitudinal mode as two ground states,
\begin{equation}
\ket{\psi_{0,xy}}=\ket{g_0}, \quad \ket{\psi_{1,z}} = \ket{g_1},
\end{equation}
we arrive at an alternative interpretation of the data in Table \ref{tab:EcoRI}. Instead of the Bell state being an energy eigenmode of the longitudinal oscillations, it may be a superposition of these states in the form 
\begin{equation} \label{entstate}
\ket{\Psi}=\rho \ket{g_0} + \nu \ket{g_1}, 
\end{equation}
where $\rho \neq 0,\,\nu \neq 0, \, \abs{\rho}^2+\abs{\nu}^2=1$. We can set the expectation value equal to the double-strand breakage energy:
\begin{align} \label{expH}
\nonumber \bra{\Psi}H\ket{\Psi}&=\abs{\rho}^2\hbar\Omega_{0,xy}/2 + \abs{\nu}^2\hbar\Omega_{1,z}/2 \\ 
&\stackrel{!}{=} 2 \varepsilon_{P-O},
\end{align}
which, when taken together with the completeness relation above for this restricted Hilbert space, gives the following probabilities for our two-qubit state:
\begin{equation}
\abs{\rho}^2 \approx 0.67, \quad \abs{\nu}^2 \approx 0.33.
\end{equation}
If the entangled state is not an energy eigenmode, but rather a coherent superposition of normal modes, then these amplitudes suggest that it is about twice as likely to be measured in $\ket{g_0}$ as in $\ket{g_1}$. To verify such entanglement, one route would be to confirm non-classical correlations between spatially separated base pairs by measuring, for example, a Bell inequality in a single-molecule experiment. Though extremely challenging for current setups, if successful, we would be able to determine the relative probability and phase between the ground states in Equation \ref{entstate}.

Such an entangled state implies more than quantum coherence alone. Before decay of the oscillatory quanta into the phosphodiester bonds, one cannot predict deterministically the outcome of the binding event at one catalytic center. There would be a coordinated likelihood (equal only when $\abs{\alpha}=\abs{\beta}=\frac{1}{\sqrt{2}}$) for the strand to be cut or not cut. However, if the entangled mode results in a strand break at one catalytic center, we can know instantaneously by the laws of quantum mechanics that the other strand is broken. What was completely random at both locations before measurement becomes, after measurement at a single site, completely certain at the other, distant site. 

Why might the zero-point mode decay symmetrically to break two bonds? The symmetry of orthodox type II recognition sequences should conserve parity during the process, thereby ensuring symmetrical bond breaks. However, this symmetry may be disturbed by various mechanisms, including enzyme modification \cite{Pingoud3, Pingoud2} and phosphorothiorate substitution \cite{Potter}, in which case it is possible that single-strand breakage might occur. Such DNA ``nicking'' is precisely the observation in several type II systems, including \textit{Eco}RV, for which the catalytic activity is reduced to one-half that of wild-type \textit{Eco}RV when mutated asymmetrically in one subunit \cite{Pingoud3}. Disruption of the \textit{Eco}RV symmetry by genetic modification disrupts parity conservation and therefore results in independent single-strand breaks. Catalytic activity is reduced due to the sub-optimal performance requirement of dissociation and re-binding to complete the second cut.  Even relatively mild asymmetric perturbations can profoundly change catalytic rate constants, as illustrated by the introduction of a guanine analogue in one strand of the \textit{Eco}RI site, whereby such constants were decreased by up to 30-fold \cite{Jen-Jacobson}. Such marked symmetry violation corresponding to the introduction of asymmetry in the base-pair ladder confirms the primacy of DNA sequence mediation in type II endonuclease catalysis.

\begin{figure}[htb]
\centering
\includegraphics[width=0.65\textwidth]{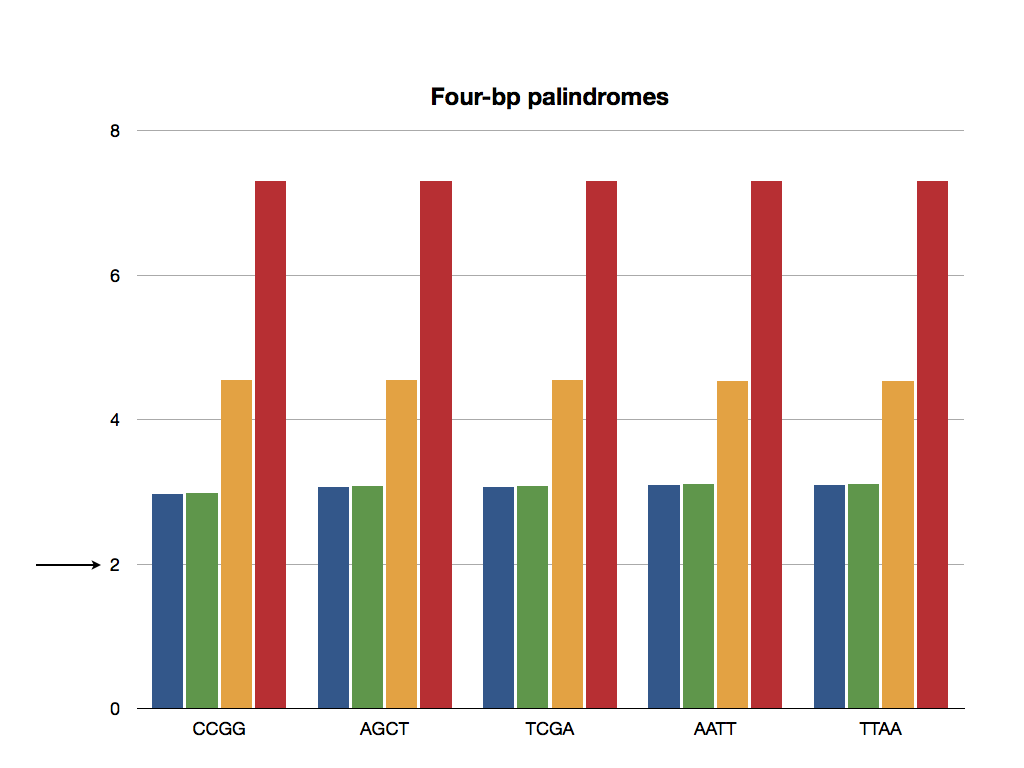}
\caption{Type II endonuclease four-bp recognition sequence zero-point modes, in units of $\varepsilon_{P-O} \simeq 0.23$ eV, with $d=3.4\rm \r{A}$. $\hbar\Omega_{0,xy}/2$ (blue), $\hbar\Omega_{1,xy}/2$ (green), $\hbar\Omega_{0,z}/2$ (gold), $\hbar\Omega_{1,z}/2$ (red) with 2$\varepsilon_{P-O}$ quanta highlighted.}
\label{fig:4bp}
\end{figure}

\begin{figure}[htb]
\centering
\includegraphics[width=0.65\textwidth]{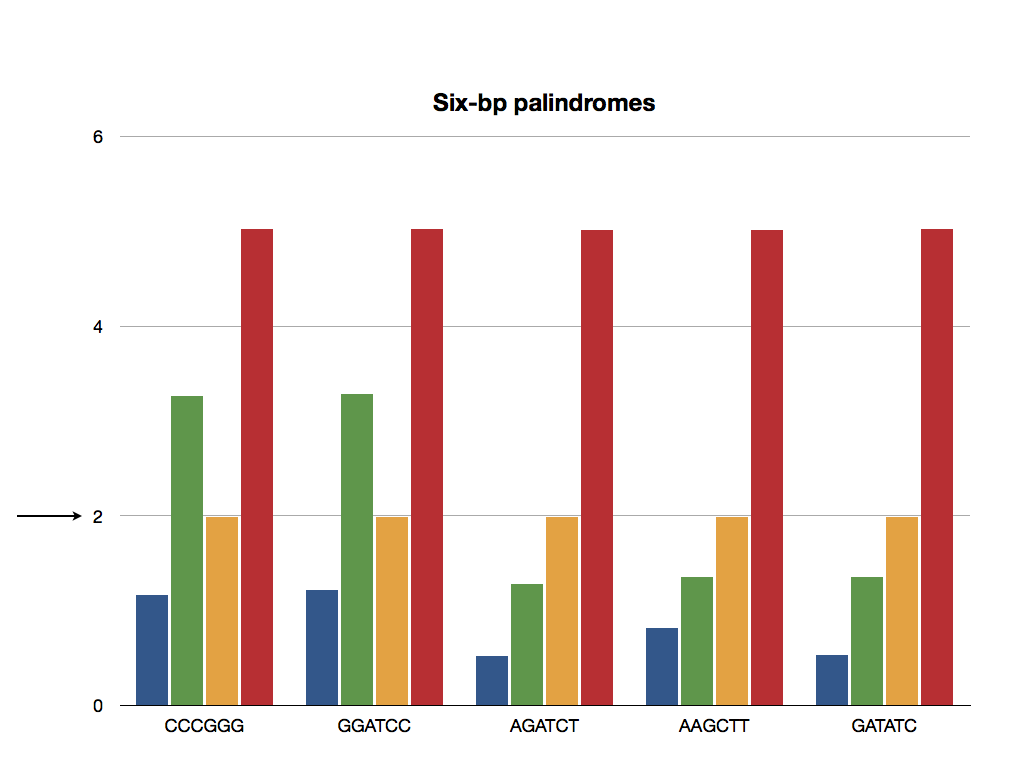}
\caption{Type II endonuclease six-bp recognition sequence zero-point modes, in units of $\varepsilon_{P-O} \simeq 0.23$ eV, with $d=3.4\rm \r{A}$. $\hbar\Omega_{0,xy}/2$ (blue), $\hbar\Omega_{1,xy}/2$ (green), $\hbar\Omega_{0,z}/2$ (gold), $\hbar\Omega_{1,z}/2$ (red) with 2$\varepsilon_{P-O}$ quanta highlighted.}
\label{fig:6bp}
\end{figure}

\begin{figure}[htb]
\centering
\includegraphics[width=0.65\textwidth]{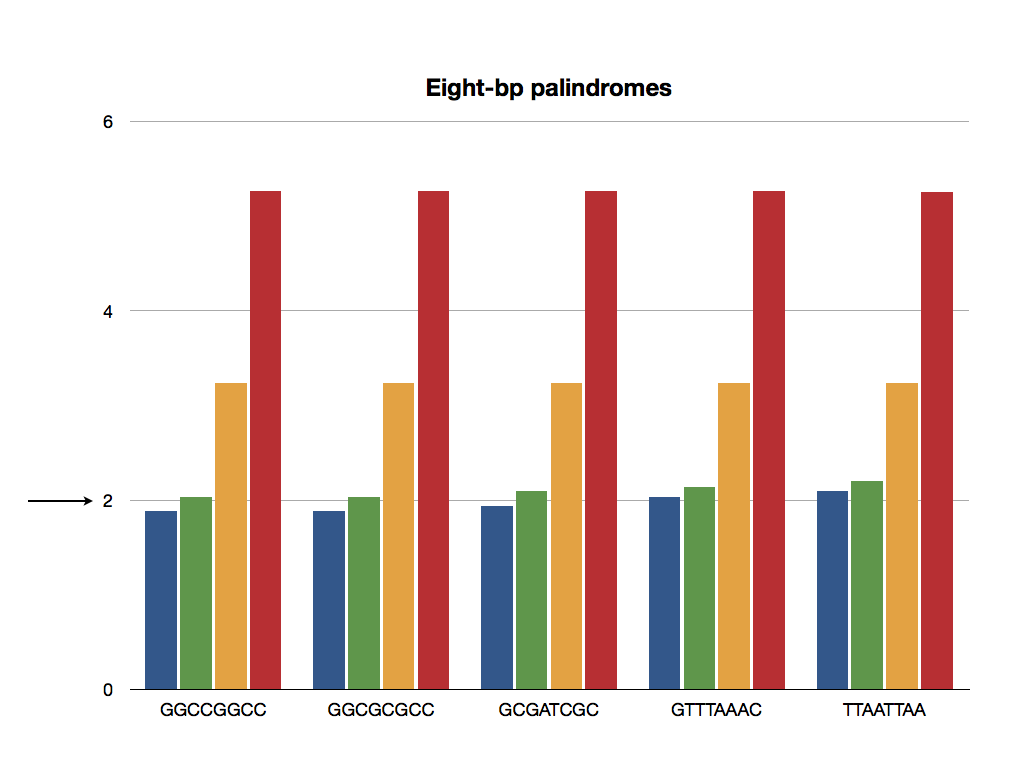}
\caption{Type II endonuclease eight-bp recognition sequence zero-point modes, in units of $\varepsilon_{P-O} \simeq 0.23$ eV, with $d=3.4\rm \r{A}$. $\hbar\Omega_{0,xy}/2$ (blue), $\hbar\Omega_{1,xy}/2$ (green), $\hbar\Omega_{0,z}/2$ (gold), $\hbar\Omega_{1,z}/2$ (red) with 2$\varepsilon_{P-O}$ quanta highlighted.}
\label{fig:8bp}
\end{figure}

Results for other type II restriction endonucleases are presented in Figures \ref{fig:4bp}, \ref{fig:6bp}, and \ref{fig:8bp}. Although the data is not exhaustive, the recognition sequences presented here \cite{REBASE} do encompass the spectrum of results, and data not shown will fall within this spectrum. The energies generally exhibit local minima at six base pairs, where the $0$th $z$-mode is most closely tuned to $2\varepsilon_{P-O}$, indicating that such sequences may have conferred an evolutionary advantage in catalytic efficiency. Indeed, the vast majority of type II enzymes recognize target sequences that are six base pairs in length. 

Figure \ref{fig:8bp} paints a slightly different picture for eight-base-pair recognition sequences. Though the $z$-modes are too large to provide the right quanta necessary for double-strand breakage, all $xy$-modes displayed are within ten percent of the threshold. It is certainly possible that a single eigenmode provides the energy for two cuts, but as in the example with \textit{Eco}RI above, it could also be a coherent superposition of normal modes that stimulates synchronized catalysis. 

The four-base-pair recognition sequences displayed in Figure \ref{fig:4bp} exhibit transverse energies that are more than $50\%$ larger than the threshold, with longitudinal energies two to four times that value. This case suggests that the recruitment of quanta for double-strand breakage likely takes a path modified from the one described by our entangled state model. These findings agree with what one might expect from comparison of type II protein sequences alone and their DNA recognition strategies, which reveal little or no similarity. 

The sequence dependence of our model is manifested most starkly in the behavior of the transverse modes for each length recognition sequence. As might be expected from Table \ref{DNAbpomega}, the transverse mode energies are largest for A:T-rich sequences (e.g., TTAA, TTAATTAA) and smallest for C:G-rich sequences (e.g., CCGG, GGCCGGCC) in the four- and eight-base-pair cases. Additional complex harmonics in the six-base-pair case (for CCCGGG and GGATCC) alter this pattern. Since $\omega_{A:T, zz}$ is within $0.05\%$ of $\omega_{C:G, zz}$, we see little or no variation in $\hbar\Omega_{0,z}/2$ or $\hbar\Omega_{1,z}/2$ within each sequence length category.

\begin{figure}[htb]
\centering
\includegraphics[width=0.99\textwidth]{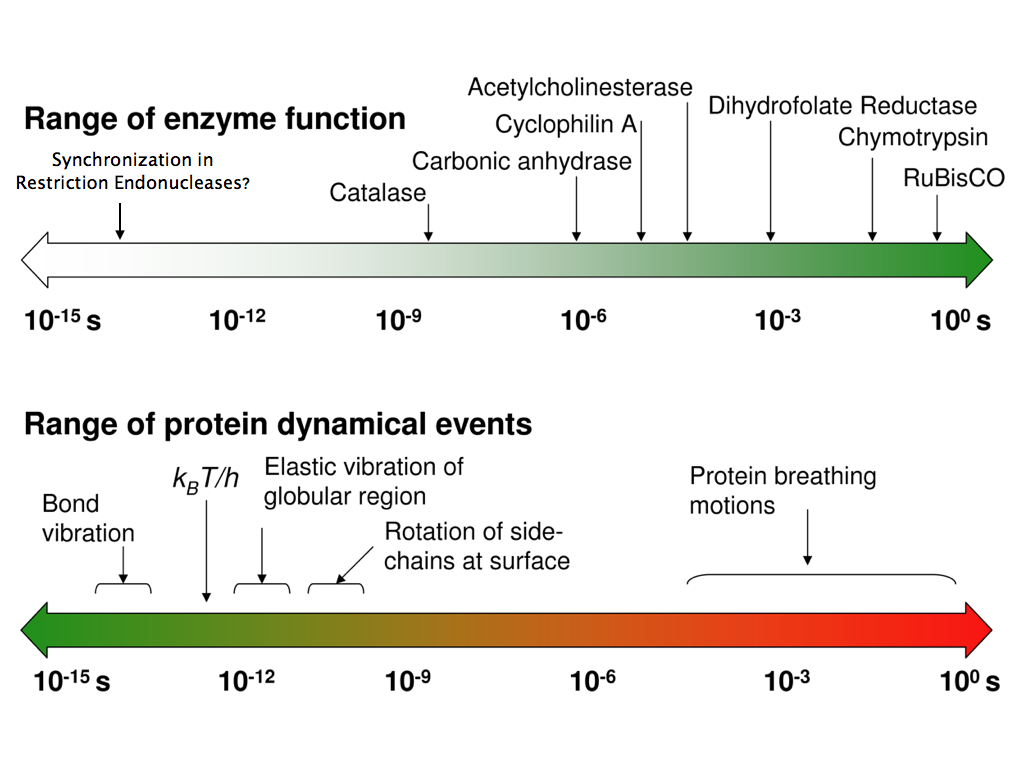}
\caption{The range of timescales involved in enzyme-catalyzed reactions and implicated in protein dynamics are similar. Quantum catalytic synchronization in type II restriction endonucleases would occur near the far left end of the spectrum, on the order of electronic oscillations. (Modified from \cite{agarwal2006enzymes}.)}
\label{fig:enz}
\end{figure}

The case against quantum entanglement mediating catalytic synchronization would be bolstered by the existence of a physically viable causal alternative---whether mechanical, electronic, or otherwise. Perhaps the strongest argument \cite{Crosstalk} suggests the formation of a ``crosstalk ring'' in \textit{Eco}RI that couples recognition to catalysis in both catalytic sites through the formation of a hydrogen bond network between select base pairs and interdigitated amino acid residues. Such a biochemical communication system would have to overcome the randomizing influence of thermal motions; consequently, the crosstalk mechanism is physically unsound on the basis of the timescale required for slow residue positioning (1\,ns) to synchronize catalytic behavior across significant distance ($20\, \rm \r{A}$). Sub-picosecond timescale electronic fluctuations, such as those described herein, have been shown to facilitate the formation of the larger-scale catalytic state across a wide range of protein energy landscapes \cite{henzler2007hierarchy, agarwal2006enzymes}. The longer, nanosecond to second timescale dynamics are usually rate-limited by the thermodynamically driven diffusion required to bind appropriate substrate---the so-called substrate turnover step---as is the case for restriction endonucleases. While changes in average geometry from enzyme-substrate complex to transition state also occur on the nanosecond to second timescale, analysis of real-time dynamic trajectories from the transition state \cite{agarwal2002network} inform us about fluctuations occurring on the femtosecond timescale, as shown in Figure \ref{fig:enz}. This crosstalk alternative thus does not explain the absolute coherence between synchronized cuts in several native \textit{Eco}RI-like systems, because we would expect thermal motions to disrupt the formation of the bond network with non-negligible probability on the longer timescale, thus producing single cuts. Instead, faster oscillations that are protected from thermal motions in the critical transition state would synchronize the catalytic centers.

From these data we may draw the tentative conclusion that many orthodox type II restriction endonucleases---recognizing target sequences of six or eight base pairs---are biologically optimized for exploiting quantum coherence in synchronized bond breaking. When an asymmetry is introduced in enzyme or substrate, it is likely that these catalysts switch to a non-synchronized form of cutting in which an intermediate nicked DNA is produced. This non-synchronized form is explained quite well by classical biochemical means such as the crosstalk mechanism proposed for \textit{Eco}RI. Structurally symmetrical DNA substrates are essential for the decay of the coherent state quanta into two locally identical phosphodiester bonds; asymmetric cuts occur due to perturbation from ideal conditions, in which quantum catalytic synchronization is traded for less efficient means. The correlation between one single-strand cut and another then need not obey the absolute coherence of quantum entanglement, as thermodynamic probabilities and traditional kinetics would apply.

The thermal de Broglie wavelength $\hbar \sqrt{2\pi/mk_BT}$ of physiological electrons is on the order of four nanometers, about twice the length of orthodox type II recognition sequences, suggesting that quantum effects would dominate in this regime. Furthermore, the probability that a subsystem immersed in a thermal bath will have a particular energy $E$ goes as $\exp(-E/k_BT)$, which is $O(10^{-8})$ for our threshold oscillations. Without accounting for normalization by the partition function, this is a reasonable figure given that under optimum conditions restriction endonucleases can scan up to $10^6$ base pairs in a single binding event. 

Strictly speaking, such decoherence-shielded systems are not in thermal equilibrium with the environment. We can estimate the decoherence time from the Heisenberg uncertainty relation
\begin{equation}
\Delta E \Delta \tau \geq \frac{\hbar}{2},
\end{equation}
which gives a lower bound on $\Delta \tau \geq (\Delta \Omega)^{-1} = O(10^{-15}) \rm \,s$. This estimate is not surprising for two reasons: We would expect synchronization to occur faster than the molecular collision time, and decoherence time represents the timescale over which phase coherence between energy eigenstates is lost, not the timescale for the system to ``become classical.'' Although biological quantum systems can be extremely fragile in terms of maintaining superpositions of energy eigenstates, they can be extremely robust in maintaining superpositions of readout (e.g., $\ket{cc}$ or $\ket{nn}$) eigenstates, which are the actual states of the system.

The reader may ask whether alternative vibrational modes of DNA might stimulate synchronized cutting. Acoustic phonons that excite strictly mechanical modes of the DNA helix have been observed by dielectrometry \cite{Edwards}, Brillouin scattering \cite{Maret, Hakim}, and Raman spectroscopy \cite{Urabe, Urabe2}. Some authors have speculated about the transport of coherent energy over distance to site-specific locations \cite{Edwards}, although possible demonstrations of biochemical effect were not proposed. Treating the double-stranded DNA as a linear spring \cite{Bustamante} with small displacements from equilibrium, we can obtain Hooke's constant $k_{H}=3k_BT/2PL$, where $P \approx 50$nm is the persistence length of DNA in physiological salt and $L$ is the length of the strand. This corresponds to an oscillation frequency of approximately $3 \times 10^9$ radians per second and an energy six orders of magnitude smaller than that of $\varepsilon_{P-O}$. Therefore the mechanical modes of vibration do not compete with the electronic dipole-dipole oscillations in the helix for the purpose of breaking phosphodiester bonds.

\section{Implications for DNA Sequence Regulation}
What should be apparent is the intricate complementarity of orthodox type II restriction endonucleases to their DNA substrates, and not simply at the biochemical sequence level. Akin to the models for olfactory sensing \cite{Brookes, Franco}, where molecular size and shape are involved but quantum effects are also exploited, electronic collective modes in the DNA helix generate oscillatory quanta that appear fine-tuned to the job of coherently cutting two phosphodiester bonds. Preparation of the decoherence-free subspaces which maintain such coherent energy transfer is initiated by orthodox type II endonucleases upon specific binding to the appropriate recognition sequences, when intimate contacts between protein and DNA force out water and ions that might disturb the delicate quantum state within the enzyme pocket. As shown in Figure \ref{Debyegraphs}, electrostatic effects are exponentially screened in this pocket outside even a fraction of the distance between base pairs, providing the proper milieu for biology to utilize quantum phenomena to enhance functionality. 

It is intriguing that quantum coherence as proposed in these biological systems would serve as a means to optimize energy transfer. In photosynthetic light-harvesting complexes \cite{Sarovar}, quantum entanglement has been implicated over picosecond timescales to optimize energy transfer between the impinging photon and the chemical reaction center. With orthodox type II endonucleases, where no ATP or other chemical energy source is required for catalysis, it would be fitting if in genome evolution nature has again chosen quantum entanglement to appropriate energy in a site-specific and symmetry-conserving manner. It is no coincidence, then, that the evolutionary best for catalytic rates occurs by what we have proposed in this paper as \textit{in vivo} quantum coherent synchronization. That this phenomenon would occur in DNA suggests the centrality and ubiquity of quantum coherent energy transfer in living systems.

What also becomes clear is the extraordinary sensitivity of these enzymes to environmental parameters. Just as complex systems exhibit behavior that cannot be predicted from the mechanics of microscopic constituents, so biology has dynamically optimized several parameters to achieve maintenance of the proposed DNA quantum state. \textit{Eco}RV incubated at ideal pH cuts both strands of the DNA in the synchronized, concerted manner discussed in this paper; in contrast, the enzyme reaction at lower pH involves sequential, independent cutting of the two strands \cite{Halford}. The difference in catalysis, which cannot be accounted for by weakened enzyme-DNA binding, has been traced to the asymmetrical binding of Mg$^{2+}$ to the \textit{Eco}RV subunits. Thus we see that the pH disturbance propagates through the buffer solution, generating a local electromagnetic environment in which the symmetry of the complex is broken. Similar sub-optimal results hold for low concentrations of MgCl$_2$.  

Disruption of symmetry and disturbance from biological optimality serve as beacons for the genesis of non-synchronized cuts. When the DNA substrate is synthesized asymmetrically by single-strand phosphorothiorate substitution, \textit{Eco}RI, \textit{Bam}HI, and \textit{Hin}dIII all cleave at a reduced rate and by a sequential process in which the DNA is converted to a nicked intermediate. Several other type II cases demonstrate isolable intermediates with single-strand scissions under sub-optimal reaction conditions, including with supercoiled DNA \cite{Supercoiling} or at low temperatures \cite{Potter}. Supercoiling compacts the helix, and both Figure \ref{subfig:dsAAAA} and Table \ref{tab:EcoRI} show that the $1/d^3$ dependence for the interaction potential may displace the zero-point energies far from the $2 \varepsilon_{P-O}$ threshhold required for synchronized catalysis. At low temperatures, both pH and enzyme binding are affected, which would change the local electromagnetic environment and introduce asymmetries in the protein-DNA complex. In contrast to the difficulty encountered in many quantum coherence measurements, biology may be able to create entangled states more effectively at physiological temperature. 

To explain the coupling between recognition and catalysis, speculative biochemical mechanisms have invoked significant structural rearrangements of the DNA substrate from its observed conformation in crystals \cite{Horton}, highlighting the inadequacy of classical models and pointing, ultimately, to the limits of causal communications. The data are consistent with our model of entangled states that bridges this knowledge gap. We submit that what has been perceived as distinct electrons communicating biochemically can now be seen as one quantum entity instantaneously collapsing into the observed state. The inability to deduce mechanistic information conclusively from crystallized snapshots in time does indeed support the inherently nonlocal role of quantum entanglement in synchronizing the catalytic centers of several type II restriction endonucleases.

\section{Discussion and Outlook}
The genome is populated by clamping, site-specific DNA-binding proteins. Palindromic inverted repeats such as those recognized by type II endonucleases are hallmarks of the RAG-mediated cutting in the immune system V(D)J joining process as well as in DNA transposon processes that are catalytically similar to HIV-1 integration \cite{Reznikoff}. The widely studied \textit{lac} repressor has been shown to recognize the palindromic symmetry of its lactose operator sequence, with enhanced binding to artificial substrates that preserve symmetry but differ significantly from naturally occurring targets \cite{Lac}. Such common phylogenetic heritage in nucleotide symmetry, possibly enabling quantum coherent energy transfer, suggests the potential widespread exploitation of decoherence-free subspaces in living systems. 

Integral to the maintenance and proliferation of genome diversity in these systems is meiotic recombination---a process initiated by DNA double-strand breaks that are catalyzed during chromosome pairing and alignment, before segregation. The molecular basis of the distribution of double-strand breaks in chromosomal populations has remained elusive. While the proteins associated with this catalytic event are many and interact as a complex system, a working model \cite{keeney2008spo11} centers on the enzyme Spo11's concerted action and implication in endonuclease-like formation of double-strand breaks. Short palindromic and symmetrical sequences are involved in the recruitment of Spo11 to the recombination site. The nature of Spo11 dimer activation is not clear but is surmised to involve a tight-binding clamp-like stabilization complex of Spo11 on the DNA that decays to form two cuts. Though admittedly a different biology, the Spo11 complex exhibits stark similarities to the quantum entanglement mechanism suggested for catalytic synchronization of several type II restriction endonucleases. We submit that the quantum model proposed herein extends a first-principles explanation of DNA double-strand break formation and addresses an outstanding problem in molecular genetics by its useful application to biological mechanisms of profound significance to genome diversity. 

Short inverted repeats have been shown to play crucial roles in DNA metabolism, but long inverted repeats are rare and can be a source of genome instability. In yeast, a perfect palindrome formed by two one-kilobase inverted repeats increases interchromosomal recombination in the adjacent region 17,000-fold \cite{Nematode}. Observation of palindromic motifs greater than two kilobases in three species of roundworm \cite{Nematode} suggests that the conserved structure in the intergenic sequence is due to selection for some function that requires the unique physical properties allowed by inverted repeats. Certainly, their occurrence in this non-protein-coding region implicates their symmetry in some aspect of DNA sequence regulation. That the palindromic symmetry is conserved---and not the biochemical sequence of the base pairs---points to a quantum mechanism similar to what we describe in this paper.

In conventional thermodynamics, the state of a macroscopic system never hinges on the outcome of a single microscopic event. However, with the advent of single-molecule detection techniques for manipulating cellular components, we can examine regimes far from this realm, where quantum uncertainty could actually determine the fate of a biological system. The discovery of quantum states in protein-DNA complexes would thus allude to the tantalizing possibility that these complexes might be prime candidates for biological quantum computation. Indeed, the evidence is mounting for such technology. Instantaneous transfer of spin coherence has been demonstrated in quantum dots at ambient temperatures \cite{Ouyang}. The biological feasibility of quantum computation in brain microtubules \cite{Hameroff} has been proposed. Nano-electromechanical systems modeled by arrays of coupled oscillators can demonstrate transport of quantum entanglement \cite{Eisert}. Development of \textit{in vitro} techniques for ultrafast pulsed lasers could allow us to exploit the quantum Zeno effect and  freeze windows into the evolution of the biological quantum state. 

First steps have been taken toward connecting the mathematical formalism of quantum computation with our model for DNA electronic vibrations. Grover's database search algorithm---the optimal algorithm for selecting a desired object from an unsorted collection of items---uses quantum superposition to reduce the number of required queries from $O(N)$ to $O(\sqrt{N})$. Intriguingly, a physical scheme for the algorithm has been proposed using coupled simple harmonic oscillators \cite{AP}, for which the simplest case solution (a single query) is $N=4$. This scheme has been implicated in efficient transport and site-specific focusing of oscillatory energy. 

Biology is characterized by macroscopic open systems in non-equilibrium conditions. Macroscopic organization of microscopic components (e.g., molecules, ions, electrons) that exhibit quantum behavior is rarely straightforward. Knowing the microscopic details of the constituent interactions and their mechanistic laws is not sufficient. Rather, as this work has shown, molecular systems must be contextualized in their local biological environments to discern appreciable quantum effects.

\section{Acknowledgements}
The authors are grateful to Tristan H\"{u}bsch, Silvina Gatica, William Southerland, and Abhijit Sarkar for sharing their valuable insights. PK was supported by an NPSC fellowship, and this work has been supported in part by NIH grants S06GM08016 and G12MD007597.


\newpage
\begin{thebibliography}{99}

\bibitem{Sarovar}
Sarovar, M., Ishizaki, A., Fleming, G. R. and Whaley, K. B. Quantum entanglement in photosynthetic light-harvesting complexes. \textit{Nature Phys.} \textbf{6,} 462--467 (2010).

\bibitem{Gauger}
Gauger, E. M., Rieper, E., Morton, J. J. L., Benjamin, S. C. and Vedral, V. Sustained quantum coherence and entanglement in the avian compass. \textit{Phys. Rev. Lett.} \textbf{106,} 040503 (2011).

\bibitem{Franco}
Franco, M. I., Turin, L., Mershin, A. and Skoulakis, E. M. C. Molecular vibration-sensing component in \textit{Drosophila melanogaster} olfaction. \textit{Proc. Natl. Acad. Sci.} \textbf{108,} 3797--3802 (2011).

\bibitem{Brookes}
Brookes, J. C., Hartoutsiou, F., Horsfield, A. P. and Stoneham, A. M. Could humans recognize odor by phonon-assisted tunneling? \textit{Phys. Rev. Lett.} \textbf{98,} 038101 (2007).

\bibitem{Modrich}
Modrich, P. Studies on sequence recognition by type II restriction and modification enzymes. \textit{CRC Crit. Rev. Biochem.} \textbf{13,} 287--323 (1982).

\bibitem{Jeltsch}
Jeltsch, A. et al. Evidence for substrate assisted catalysis in the DNA cleavage of several restriction endonucleases. \textit{Gene} \textbf{157,} 157--162 (1995).

\bibitem{Jen-Jacobson}
Jen-Jacobson, L. Protein-DNA recognition complexes: conservation of structure and binding energy in the transition state. \textit{Biopoly.} \textbf{44,} 153--180 (1997).

\bibitem{Pingoud1}
Pingoud, A. and Jeltsch, A. Structure and function of type II restriction endonucleases. \textit{Nucleic Acids Res.} \textbf{29,} 3705--3727 (2001).

\bibitem{PingoudV}
Pingoud, V. et al.  Evolutionary relationship between different subgroups of restriction endonucleases. \textit{J. Biol. Chem.} \textbf{277,} 14306--14314 (2002).

\bibitem{Roberts}
Roberts, R. J. et al. A nomenclature for restriction enzymes, DNA methyltransferases, homing endonucleases and their genes. \textit{Nucleic Acids Res.} \textbf{31,} 1805--1812 (2003).

\bibitem{PingoudV2}
Pingoud, V. et al.  Specificity changes in the evolution of type II restriction endonucleases: A biochemical and bioinformatic analysis of restriction enzymes that recognize unrelated sequences. \textit{J. Biol. Chem.} \textbf{280,} 4289--4298 (2005).

\bibitem{Type2}
Pingoud, A., Fuxreiter, M., Pingoud, V. and Wende, W. Type II restriction endonucleases: structure and mechanism. \textit{Cell. Mol. Life Sci.} \textbf{62,} 685--707 (2005).

\bibitem{Pingoud3}
Stahl, F., Wende, W., Wenz, C., Jeltsch, A. and Pingoud, A. Intra- vs intersubunit communication in the homodimeric restriction enzyme EcoRV: Thr 37 and Lys 38 involved in indirect readout are only important for the catalytic activity of their own subunit. \textit{Biochemistry} \textbf{37,} 5682--5688 (1998).

\bibitem{Pingoud2}
Stahl, F., Wende, W., Jeltsch, A. and Pingoud, A. Introduction of asymmetry in the naturally symmetric restriction endonuclease EcoRV to investigate intersubunit communication in the homodimeric protein. \textit{Proc. Natl. Acad. Sci.} \textbf{93,} 6175--6180 (1996).

\bibitem{Halford}
Halford, S. E. and Goodall, A. J. Modes of cleavage by the EcoRV restriction endonuclease. \textit{Biochemistry} \textbf{27,} 1771--1777 (1988).

\bibitem{Maxwell}
Maxwell, A., and Halford, S. E. The SalGI restriction endonuclease: mechanism of DNA cleavage. \textit{Biochem. J.} \textbf{203,} 85--92 (1982).

\bibitem{EPR}
Einstein, A., Podolsky, B. and Rosen, N. Can quantum-mechanical description of physical reality be considered complete? \textit{Phys. Rev.} \textbf{47,} 777--780 (1935).

\bibitem{Horton}
Horton, N. C., Newberry, K. J. and Perona, J. J. Metal ion-mediated substrate-assisted catalysis in type II restriction endonucleases. \textit{Proc. Natl. Acad. Sci.} \textbf{95,} 13489--13494 (1998).

\bibitem{chin2013role}
Chin, A. W., Prior, J., Rosenbach, R., Caycedo-Soler, F., Huelga, S. F. and Plenio, M. B. The role of non-equilibrium vibrational structures in electronic  coherence and recoherence in pigment-protein complexes.\textit{Nature Phys.} \textbf{9,} 113--118 (2013).

\bibitem{whaley2011quantum}
Whaley, K. B., Sarovar, M. and Ishizaki, A. Quantum entanglement phenomena in photosynthetic light harvesting complexes. \textit{Procedia Chem.} \textbf{3,} 152--164 (2011).

\bibitem{Water1}
Sidorova, N. Y. and Rau, D. C. Differences in water release for the binding of EcoRI to specific and nonspecific DNA sequences. \textit{Proc. Natl. Acad. Sci.} \textbf{93,} 12272--12277 (1996).

\bibitem{Water2}
Sidorova, N. Y., Muradymov, S. and Rau, D. C. Differences in hydration coupled to specific and nonspecific competitive binding and to specific DNA binding of the restriction endonuclease BamHI. \textit{J. Biol. Chem.} \textbf{281,} 35656--35666 (2006). 

\bibitem{Water3}
Sidorova, N. Y., Muradymov, S. and Rau, D. C. Solution parameters modulating DNA binding specificity of the restriction endonuclease EcoRV. \textit{Fed. Eur. Biochem. Soc. J.} \textbf{278,} 2713--2727 (2011).

\bibitem{Cocco}
Cocco, S. and Monasson, R. Theoretical study of collective modes in DNA at ambient temperature. \textit{arXiv:cond-mat/9911008v1 [cond-mat.stat-mech]} (1999).

\bibitem{Blinov}
Blinov, V. N. and Golo, V. L. Acoustic spectroscopy of DNA in the gigahertz range. \textit{Phys. Rev. E} \textbf{83,} 021904 (2011).

\bibitem{Woolard}
Woolard, D. L. et al. Submillimeter-wave phonon modes in DNA macromolecules. \textit{Phys. Rev. E} \textbf{65,} 051903 (2002).

\bibitem{Rieper}
Rieper, E., Anders, J. and Vedral, V. Quantum entanglement between the electron clouds of nucleic acids in DNA. \textit{arXiv:1006.4053v2 [quant-ph]} (2011).

\bibitem{Fock-Dirac}
McWeeny, R. Perturbation theory for the Fock-Dirac density matrix. \textit{Phys. Rev.} \textbf{126,} 1028--1034 (1962).

\bibitem{CNDO}
Papadopoulos, M. G. and Waite, J. The polarisability and second hyperpolarisability of some biomolecules. \textit{J. Mol. Struct. (Theochem)} \textbf{170,} 189--196 (1988).

\bibitem{Basch}
Basch, H., Garmer, D. R., Jasien, P. G., Krauss, M. and Stevens, W. J. Electrical properties of nucleic acid bases.\textit{Chem. Phys. Lett.} \textbf{163,} 514--522 (1989).

\bibitem{Cohen-Tannoudji}
Cohen-Tannoudji, C., Diu, B. and Laloe, F. \textit{Quantum Mechanics, Vol. 2} Appendix I (Wiley, New York, 1977).

\bibitem{BriggsEisfeld}
Briggs, J. S. and Eisfeld, A. Coherent quantum states from classical oscillator amplitudes. \textit{Phys. Rev. A} \textbf{85,} 052111 (2012).

\bibitem{Phospho}
Dickson, K. S., Burns, C. M. and Richardson, J. P. Determination of the free-energy change for repair of a DNA phosphodiester bond. \textit{J. Biol. Chem.} \textbf{275,} 15828--15831 (2000).

\bibitem{Potter}
Potter, B. V. L. and Eckstein, F.  Cleavage of phosphorothioate-substituted DNA by restriction endonucleases. \textit{J. Biol. Chem.} \textbf{259,} 14243--14248 (1984).

\bibitem{Crosstalk}
Kurpiewski, M. R. et al. Mechanisms of coupling between DNA recognition specificity and catalysis in EcoRI endonuclease. \textit{Structure} \textbf{12,} 1775--1788 (2004).

\bibitem{Soliton}
Satari\'{c}, M. V., Tuszy\'{n}ski, J. A. and \u{Z}akula, R. B. Kinklike excitations as an energy-transfer mechanism in microtubules. \textit{Phys. Rev. E} \textbf{48,} 589--597 (1993).

\bibitem{REBASE}
Roberts, R. J., Vincze, T., Posfai, J. and Macelis, D. REBASE---a database for DNA restriction and modification: enzymes, genes and genomes. \textit{Nucleic Acids Res.} \textbf{38,} D234--D236 (2010). 

\bibitem{henzler2007hierarchy}
Henzler-Wildman, K. A., Lei, M., Thai, V., Kerns, S. J., Karplus, M. and Kern, D. A hierarchy of timescales in protein dynamics is linked to enzyme catalysis. \textit{Nature} \textbf{450,} 913--916 (2007).

\bibitem{agarwal2006enzymes}
Agarwal, P. K. Enzymes: An integrated view of structure, dynamics and function. \textit{Microb. Cell Fact.} \textbf{5,} 2 (2006).

\bibitem{agarwal2002network}
Agarwal, P. K., Billeter, S. R., Rajagopalan, P. T. R., Benkovic, S. J. and Hammes-Schiffer, S. Network of coupled promoting motions in enzyme catalysis. \textit{Proc. Natl. Acad. Sci.} \textbf{99,} 2794--2799 (2002).

\bibitem{Edwards}
Edwards, G. S., Davis, C. C., Saffer, J. D. and Swicord, M. L. Microwave-field-driven acoustic modes in DNA. \textit{Biophys. J.} \textbf{47,} 799--807 (1985).

\bibitem{Maret}
Maret, G., Oldenbourg, R., Winterling, G., Dransfeld, K. and Rupprecht, A. Velocity of high frequency sound waves in oriented DNA fibres and films determined by Brillouin scattering. \textit{Colloid and Polymer Sci.} \textbf{257,} 1017--1020 (1979).

\bibitem{Hakim}
Hakim, M. B., Lindsay, S. M. and Powell, J. The speed of sound in DNA. \textit{Biopolymers} \textbf{23,} 1185--1192 (1984).

\bibitem{Urabe}
Urabe, H. and Tominaga, Y. Low-lying collective modes of DNA double helix by Raman spectroscopy. \textit{Biopolymers} \textbf{21,} 2477--2481 (1982).

\bibitem{Urabe2}
Urabe, H., Tominaga, Y. and Kubota, K. Experimental evidence of collective vibrations in DNA double helix (Raman spectroscopy). \textit{J. Chem. Phys.} \textbf{78,} 5937--5939 (1983).

\bibitem{Bustamante}
Bustamante, C., Smith, S. B., Liphardt, J. and Smith, D. Single-molecule studies of DNA mechanics. \textit{Curr. Opin. Struct. Biol.} \textbf{10,} 279--285 (2000).

\bibitem{Supercoiling}
Ruben, G. et al. Relaxed circular SV40 DNA as cleavage intermediate of two restriction endonucleases. \textit{Nucleic Acids Res.} \textbf{4,} 1803--1813 (1977).

\bibitem{Reznikoff}
Reznikoff, W. S. MicroReview: Tn5 as a model for understanding DNA transposition. \textit{Molec. Microbiol.} \textbf{47,} 1199--1206 (2003).

\bibitem{Lac}
Sadler, J. R., Sasmor, H. and Betz, J. L. A perfectly symmetric lac operator binds the lac repressor very tightly. \textit{Proc. Natl. Acad. Sci.} \textbf{80,} 6785--6789 (1983).

\bibitem{keeney2008spo11}
Keeney, S. Spo11 and the formation of DNA double-strand breaks in meiosis. In \textit{Recombination and Meiosis} 81--123 (2008).

\bibitem{Nematode}
Zhao, G., Chang, K. Y., Varley, K. and Stormo, G. D. Evidence for active maintenance of inverted repeat structures identified by a comparative genomic approach. \textit{Pub. Lib. Sci. ONE} \textbf{2,} e262 (2007).

\bibitem{Ouyang}
Ouyang, M. and Awschalom, D. D. Coherent spin transfer between molecularly bridged quantum dots. \textit{Science} \textbf{22,} 1074--1078 (2003).

\bibitem{Hameroff}
Hagan, S., Hameroff, S. R. and Tuszynski, J. A. Quantum computation in brain microtubules: Decoherence and biological feasibility. \textit{Phys. Rev. E} \textbf{65,} 061901 (2002).

\bibitem{Eisert}
Eisert, J., Plenio, M. B., Bose, S. and Hartley, J. Towards quantum entanglement in nanoelectromechanical devices. \textit{Phys. Rev. Lett.} \textbf{93,} 190402 (2004).

\bibitem{AP}
Patel, A. A. and Patel, A. D. A coupled oscillator model for Grover's quantum database search algorithm. \textit{arXiv:0711.4733v1 [physics.gen-ph]} (2007).


\end{thebibliography}
\end{document}